\documentclass[floatfix,aps,twocolumn,superscriptaddress,prl]{revtex4-2}

\usepackage{graphicx}
\usepackage{dcolumn}
\usepackage{bm}
\usepackage[T1]{fontenc}
\usepackage[latin9]{inputenc}
\usepackage{textcomp}
\usepackage{amsmath}
\usepackage{mathrsfs}
\usepackage{graphicx}
\usepackage{amssymb}
\usepackage{siunitx}
\usepackage{xcolor}
\usepackage{float}
\usepackage[colorlinks,citecolor=blue,urlcolor=blue]{hyperref}
\usepackage{MnSymbol}

\newcommand{\paolo}[1]{\textcolor{blue}{#1}}

\makeatletter
\def\maketitle{
\@author@finish
\title@column\titleblock@produce
\suppressfloats[t]}
\makeatother

\begin{document}

\title{Quantum impurities in finite-temperature Bose gases: \\Detecting vortex proliferation across the BKT and BEC transitions}

\author{Paolo Comaron}
\email{paolo.comaron@nanotec.cnr.it}
\address{CNR NANOTEC, Institute of Nanotechnology, Via Monteroni, 73100 Lecce, Italy}
\address{Department of Physics and Astronomy, University College London,
Gower Street, London, WC1E 6BT, United Kingdom} 

\author{Nathan Goldman}
\affiliation{CENOLI,
Universit\'e Libre de Bruxelles, CP 231, Campus Plaine, B-1050 Brussels, Belgium}
\affiliation{Laboratoire Kastler Brossel, Coll\`ege de France, CNRS, ENS-Universit\'e PSL, Sorbonne Universit\'e, 11 Place Marcelin Berthelot, 75005 Paris, France
}

\author{Atac Imamoglu}
\affiliation{Institute of Quantum Electronics ETH Zurich, CH-8093 Zurich, Switzerland
}

\author{Ivan Amelio}
\email{ivan.amelio@ulb.be}
\affiliation{CENOLI,
Universit\'e Libre de Bruxelles, CP 231, Campus Plaine, B-1050 Brussels, Belgium}

\begin{abstract}
Detecting vortices in neutral superfluids represents an outstanding experimental challenge. 
Using stochastic classical-field methods,
we 
theoretically show 
that a quantum impurity repulsively coupled to a
weakly-interacting Bose gas at finite temperature
carries direct spectroscopic signatures of  vortex proliferation. In two dimensions, we find that 
a low-energy (attractive) branch in the excitation spectrum becomes prominent when the temperature is tuned  across the Berezinskii-Kosterlitz-Thouless (BKT) transition. We explain this red-shifted resonance as originating from the binding of the impurity to vortices, where the bosons density (and hence, the repulsive Hartree energy) is reduced.
This mechanism could be exploited to spectroscopically estimate the BKT transition in excitonic insulators.
In contrast, in three dimensions, the impurity spectra reflect the presence of vortex rings well below the condensation temperature, and herald the presence of a thermal gas above the Bose-Einstein condensation transition. 
Importantly,
we  expect our results to have  impact on the understanding of Bose-polaron formation at finite temperatures.
\end{abstract}

\maketitle

\textit{Introduction.}
A mobile impurity immersed in a quantum background,  commonly referred to as  {\em polaron}, provides a paradigmatic setting of quantum many-bodyphysics~\cite{rath2013field,massignan2014,scazza2022,parish2023fermi,grusdt2024impuritiespolaronsbosonicquantum}.  
Studying polarons  is relevant to understanding  fundamental questions about transport and the emergence of quasi-particles~\cite{landaupekar1948}, addressing the physics of imbalanced mixtures~\cite{chevy2006universal}, 
controlling effective particle-particle interactions~\cite{baroni2023mediated},
and developing sensors for correlated many-body quantum states.

An open and challenging problem in the field concerns the fate of polarons when immersed in a finite temperature Bose gas.
A few recent studies have relied on 
perturbative calculations~\cite{Levinsen2017,Pastukhov_2018}, 
time dependent Hartree-Fock-Bogoliubov approach~\cite{Boudjema2014},
diagrammatic ladder resummation~\cite{guenther2018},
 variational methods~\cite{field2020fate,dzsotjan2020dynamical},
and a generalized functional determinant approach for the case of non-interacting bosons~\cite{drescher2024bosonicfunctionaldeterminantapproach}, while
Quantum Monte Carlo simulations have been performed in Ref.~\cite{pascual2021quasiparticle}.
In cold atoms, Bose polarons have been first observed in injection radio-frequency spectroscopy experiments~\cite{hu2016bosepolarons,jorgensen2016},
while 
ejection spectroscopy measurements close to the impurity-boson Feschbach resonance unveiled 
the role of temperature in determining the breakdown of quasi-particles in the critical regime of the mixture~\cite{Yan2020bosepolaroncriticality}.

In the context of transition-metal dichalcogenide (TMD) heterostructures, polaron spectroscopy consists in the resonant optical injection of excitons in  
the sample~\cite{sidler2017fermi,courtade2017charged}. The exciton acts as a sensor, and the quantum phase of the heterostructure reflects itself in the optical spectrum.
Several experimental and theoretical works have demonstrated the sensitivity of polaron spectroscopy to the level of doping~\cite{sidler2017fermi,courtade2017charged},  the breaking of translational symmetry~\cite{Regan2020mott,Xu2020correlated,smolenski2021signatures, Zhou2021bilayer, Li2021imaging, shimazaki2021optical,
kiper2024confinedtrionsmottwignerstates,amelio2024edpolaron},  the onset of insulating phases~\cite{Xu2020correlated,shimazaki2020strongly,colussi2023lattice,
santiagogarcia2024lattice,amelio2024polaronformationinsulatorskey,alhyder2024latticebosepolaronsstrong}, and to the pairing of electrons and holes into excitonic insulators (EXI)~\cite{amelio2023polaron,amelio2023two-dimensional,amelio2024edpolaron,qi2023electrically}, recently observed in bilayers~\cite{ma2021strongly,gu2022dipolar,Sun2021evidence,qi2023perfect,nguyen2023degenerate}.
In particular, in the context of the EXI, it is not clear whether the impurity, which interacts via short-range density-density interactions and can be used to probe pairing, would also be sensitive to the presence of superfluidity, a manifestation of quasi long-range order, and to the Berezinskii-Kostelitz-Thouless (BKT) transition~\cite{Kosterlitz1973}.

In this Letter, we study the injection spectra of an impurity repulsively coupled to a bath of weakly-interacting bosons at finite temperatures, both in two (2D) and three (3D) spatial dimensions. 
We model the Bose gas via the stochastic projected  Gross-Pitaevskii equation (SPGPE)\cite{Stoof1999,Gardiner_2002,Gardiner_2003,Blakie2008,Bradley2008}, and consider the full Schr\"odinger equation for the impurity, which is injected at zero momentum. 
When the temperature is increased, 
a low-energy spectral line emerges
 when vortices and density holes appear in the bath. 
In 2D, our results suggest that polaron spectroscopy can be used to indirectly probe the BKT transition, which is associated with the proliferation of unbound vortices.
In a similar spirit, it was recently demonstrated that indirect signatures of a topological insulator -- a phase of matter characterized by a non-local topological invariant -- can be obtained through  
 polaron spectroscopy performed locally on its edge~\cite{vashisht2024chiralpolaronformationedge}.
Also,  it was recently predicted that, for superconductors, local magnetometry provides a sensitive probe of the phase gradient of the quasi-condensate, and can provide evidence for the BKT transition~\cite{Curtis_2024}.   
In 3D, instead, we find that vortex rings proliferate well below the critical temperature of the Bose-Einstein condensate (BEC),  while at the transition the vanishing of the condensate fraction results in the disappearance of the repulsive branch.

Interestingly, detecting vortices through classical microparticles has been exploited in the context of Helium~\cite{Bewley2006,Bewley2008}. Our approach can thus be viewed as a natural generalization of this method to the realm of quantum impurities. While our proposed scheme can be directly implemented in cold-atom setups, our results also point towards the possibility of indirectly probing the BKT transition in the EXIs hosted in 2D materials.
The paper is structured as follows:
we first define the model and impurity spectra, review the SPGPE, and then discuss our results in 2D and 3D. Finally, future prospects are discussed.

~

\textit{Model and methods.}
Restricting  for simplicity to contact interactions and setting $\hbar=1$, an impurity in a weakly-interacting Bose gas can be described by the Hamiltonian
 \begin{multline}
     \hat{H}
     =
     \int d\mathbf{x} \ \left\{ \hat{\psi}^\dagger \left(-\frac{1}{2m}\nabla^2 - \mu \right)  \hat{\psi}
     - \frac{1}{2M}\hat{\Psi}^\dagger \nabla^2   \hat{\Psi}
     + 
     \right. 
     \\
     \left. 
     + \frac{g_\mathrm{BB}}{2} \hat{\psi}^\dagger \hat{\psi}^\dagger \hat{\psi} \hat{\psi}
     + g_\mathrm{BI} \hat{\psi}^\dagger \hat{\psi} \hat{\Psi}^\dagger \hat{\Psi} 
     \right\},
 \end{multline}
 where
$\hat{\psi}^\dagger(\mathbf{x})$ creates a boson at position $\mathbf{x}$ in a $d$-dimensional box of  volume $V$, and $\hat{\Psi}^\dagger(\mathbf{x})$
is the impurity creation operator (the quantum statistics of the impurity is irrelavant in this work).
The corresponding masses are $m$ and $M$,
$\mu$ is the chemical potential of the Bose gas, and $g_\mathrm{BB}$ and $g_\mathrm{BI}$ are the strength of the repulsive boson-boson and boson-impurity interactions, respectively.

In this work, we consider the injection  spectra of the impurity within the Bose gas at finite temperature.
The main observable is provided by the overlap function
\begin{equation}
    S(t) = {\rm Tr} \left[ 
    \hat{\Psi}_{\mathbf{q}=0}(t)
     \hat{\Psi}^\dagger_{\mathbf{q}=0}(0)
    \hat{\rho}(0) \right],
    \label{eq:overlap}
\end{equation}
where $\hat{\Psi}^\dagger_{\mathbf{q}=0}(t)$ is the creation operator of the impurity at zero momentum and time $t$,
and $\hat{\rho}(0)$ represents the thermal equilibrium density matrix of the bath times the impurity vacuum.
The frequency-resolved spectrum is 
$A(\omega) = -2{\rm Im} S(\omega)$,
obtained by Fourier transforming $S(t)$.
The choice of this observable is also motivated by TMD experiments, where the exciton  is optically injected with negligible momentum, with respect to the electronic scales.
In an ultracold atomic gas, one would initially prepare the impurity separately from the bath, and then quench the interactions \cite{knap2012}.

\begin{figure}[t]
\centering
\includegraphics[width=\linewidth]{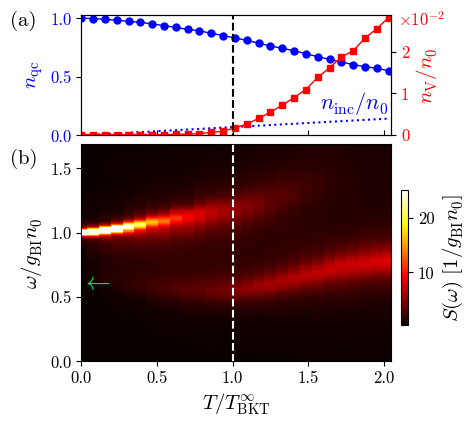} 
\caption{
\textbf{BKT phase diagrams and 2D injection spectra.}
(a) For the 2D Bose gas without impurity, we plot
the quasi-condensate fraction $n_{\rm qc}$ (blue circles, left axis),
the incoherent density $n_\mathrm{inc}$ (blue dots) and the density of vortices (red squares, right axis)
as a function of temperature. 
Errorbars are smaller than the marker size.
(b)
The injection spectrum $S(\omega)$ of the impurity is plotted as a function of $T$, in the equal mass $M=m$ case. The green arrow indicates the binding energy of the impurity in an isolated vortex at $T=0$. 
The attractive branch $S(\omega)$ displays a minimum close to the BKT transition, where vortices starts to proliferate. 
The colorcode is saturated and the lines slightly broadened, since at vanishing temperature $S(\omega)$ becomes a $\delta$-peak.  
}
\label{fig:polaron_spectra}
\end{figure}

In ultracold gases, it is established that classical-field methods, such as the GPE equation and its
generalizations, provide an excellent description of the weakly-interacting Bose gas, both at zero and finite temperature, at equilibrium or in the dynamics~\cite{Davis2001,Blakie2008,Proukakis2008,Cockburn2009,Liu2018dynamical,comaron2019quench,2021Groszek}.
In particular, the 
\paolo{SPGPE} has been recently used to study the vortex formation dynamics  across the BEC~\cite{Liu2018dynamical} and BKT~\cite{comaron2019quench} phase transitions, yielding results that are consistent with the theoretical scaling arguments.

Building on this framework, we assume that the bath consists of $N \gg 1$ bosonic atoms described at the mean-field level,
while we treat the impurity atom using its full Schr\"odinger equation.
Let us denote by $\psi$ the semi-classical field that represents the Bose gas,
and by $\Psi$ the quantum wave-function of the impurity.
In the SPGPE scheme, the equations of motion read \cite{Stoof1999,Gardiner_2002,Gardiner_2003,Blakie2008,Bradley2008}
\begin{equation}
    i\partial_t \psi = 
    \mathcal{P} (1 - i\gamma)\left\{
- \frac{1}{2m} \nabla^2
+ g_\mathrm{BB}|\psi|^2 
- \mu
    \right\} \psi + \eta,
    \label{eq:SPGPE_psi}
\end{equation}
\begin{equation}
    i\partial_t \Psi = 
    \left\{
- \frac{1}{2M} \nabla^2
+ g_\mathrm{BI} n_\mathrm{B}
    \right\} \Psi.
     \label{eq:SPGPE_Psi}
\end{equation} 
As in the traditional SPGPE framework, the Bose gas 
encompasses contributions from all highly occupied single-particle modes of the system, up to a certain cutoff in the single-particle energy spectrum.
The gas is considered to exchange energy and particles with a thermal reservoir at temperature $T$ and chemical potential $\mu$. 
The projection operator, implemented in momentum space by
$\mathcal{P} = \theta(\varepsilon_\mathrm{cut} - \frac{\mathbf{k}^2}{2m})$, prevents population transfer beyond the selected subset of single-particle modes ---the coherent region--- up to a cutoff energy
$\varepsilon_\mathrm{cut} = \mu + T\log2$, in units where the Boltzmann constant is one. 
Assuming a Bose-Einstein distribution in the occupation number spectrum, the cut-off is chosen to allow for a mean occupation of the last included mode to be of order $\sim 1$~\cite{Blakie2008}. 
The white noise $\eta$, added
and projected in the coherent region, has a variance dictated by the fluctuation-dissipation theorem,
$\langle \eta^*(\mathbf{x},t) \eta(\mathbf{x}',t') \rangle = 2 \gamma T \delta(t-t')\delta(\mathbf{x}-\mathbf{x}')$. 
The effective dissipation rate $\gamma$ regulates the coupling between the system and the bath, and can be predicted in simulations close to equilibrium~\cite{Blakie2008,Bradley2008,Rooney2013,Liu2018dynamical}.

The Bose gas density is $n_\mathrm{B}(\mathbf{x},t) = |\psi(\mathbf{x},t)|^2 + n_\mathrm{inc}$, with 
\begin{equation}
    n_\mathrm{inc} 
    =
    \int
    \frac{d\mathbf{k}}{(2\pi)^d}
    \frac{\theta(|\mathbf{k}|-k_\mathrm{cut})}{e^{{\mathbf{k}^2}/{2mT}}-1},
\end{equation}
the
incoherent part,
which
keeps into account the occupation of the bosonic modes above the cutoff blue$\varepsilon_\mathrm{cut} = k_{cut}^2/(2m)$.
In 2D, this translates into
$
    n_\mathrm{inc} 
    =
    \log2 \frac{m}{2\pi} 
     T
$,
and to
$
    n_\mathrm{inc} 
      \simeq 0.059
      (m T)^{3/2}$ in 3D.
      
In principle, a term $g_\mathrm{BI} |\Psi|^2 \psi$ should appear in Eq.~(\ref{eq:SPGPE_psi}). However, the bosonic field intensity $|\psi|^2$ is of order $N/V$, while the impurity wavefunction is normalized, which means that $|\Psi|^2$ is of order $1/V$, which goes to zero in the thermodynamic limit. In other words, when performing linear polaron spectroscopy on a large sample, the recoil of the quasi-condensate due to the impurity can be neglected, since a single delocalized impurity cannot impact the many-body (quasi-)condensate.
This assumption clearly breaks down for extremely strong impurity-boson repulsion  $g_\mathrm{BI} \sim N g_\mathrm{BB}$, where self-localization phenomena may emerge~\cite{cucchietti2006}. Also, the attractive case $g_\mathrm{BI} < 0$ may be beyond the reach of our method, since a bound state of the impurity with a single boson can play a major role in this case~\footnote{In particular, at $T=0$, one would obtain a mean-field shifted attractive line, and no repulsive branch. See Ref.~\cite{SM}.}.
We mention that the GPE approach has been recently applied to model the
strong deformation of the bosonic bath by the impurity
at zero temperature~\cite{drescher2020theory,guenther2021,yegovstsev2024exact}.

Within the SPGPE framework, 
the impurity is described by a Schr\"odinger equation with a stochastic time-dependent external potential. 
Then, the overlap of Eq.~(\ref{eq:overlap})
reduces to
    $S(t) 
   =
    \llangle \Psi_{\mathbf{q}=0}(t)
    \rrangle$, where the double brackets 
$\llangle ...
    \rrangle$ denote stochastic averages.
    Here,
    $\Psi(\mathbf{x},t)$ denotes
    the impurity wavefunction
    at time $t$, obtained from the Schr\"odinger equation (\ref{eq:SPGPE_Psi}) with uniform initial condition 
    $\Psi(\mathbf{x},0)=\frac{1}{\sqrt{V}}$.
Numerically, 
for each stochastic realization, the bath is equilibrated by evolving a uniform initial state $\psi$ via the SPGPE for a duration 
$t_\mathrm{eq}$
until
$t=0$, at  which time the impurity is initialized at momentum $\mathbf{q}=0$, and its evolution recorded.
 This procedure is averaged over $M_{\rm stat}$ stochastic realizations.
The chemical potential is adjusted with temperature so as to fix the average bosonic density $\llangle n_\mathrm{B} \rrangle =
n_0$, with $n_0$
independent of $T$.

~

\begin{figure}[t]
\centering
\includegraphics[width=0.99\linewidth]{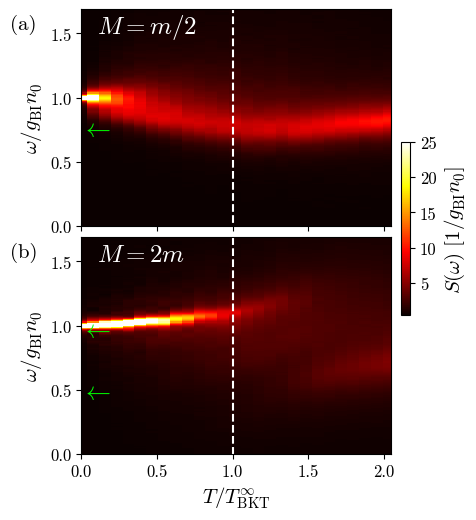} 
\caption{ 
\textbf{Injection spectra for impurities of varying masses.}
Injection spectra as a function of the  temperature, analogous to Fig.~\ref{fig:polaron_spectra}(b), are reported for a lighter $M=m/2$  and a  heavier $M=2m$ impurity, on the left and right panels, respectively. Panel (b) features two green arrows, indicating the binding energies of the two radially symmetric bound states in an ideal vortex.
}
\label{fig:masses}
\end{figure}

\textit{Numerical results in 2D.}
As it is well known, the Mermin-Wagner theorem prevents the existence of a local order parameter for the 2D Bose gas at any finite temperature.
Nonetheless, at low temperature, the field-field correlation function decays only algebraically 
$
\llangle 
{\psi}^*(\mathbf{x})
{\psi}(\mathbf{x}')
\rrangle
\propto |\mathbf{x}-\mathbf{x}'|^{-\alpha}
$,
with a critical exponent $\alpha(T)$, and  a fraction $\rho_S$ of the bosons display superfluid behavior. For a weakly-interacting gas, the BKT transition occurs at the temperature~\cite{Prokofev2002}
\begin{equation}
    T^{\infty}_{\mathrm{BKT}}
    =
    \frac{2\pi n_0}{m \log(C/m g_\mathrm{BB})},   
\end{equation}
with $C \simeq 380$; 
$\rho_S$ is found to drop to zero at the transition, 
and the critical exponent is $\alpha(T^{\infty}_{\mathrm{BKT}})=1/4$. The transition is of topological nature, and can be described in terms of a proliferation of unbound vortex-antivortex pairs.
On the other hand, 
the so-called quasi-condensate fraction
$n_{\rm qc} = \sqrt{2 \llangle |\psi|^2 \rrangle^2 - \llangle |\psi|^4 \rrangle} / \llangle |\psi|^2 \rrangle$
is smooth at threshold, since it is a local quantity. (We mention that a previous theoretical attempt to study polaron formation across the BKT transition assumed a jump in the local order parameter~\cite{alhyder2022}.)
 
This BKT physics 
is perfectly recovered at the equilibration 
step of our numerical simulations,
as illustrated in  Fig.~\ref{fig:polaron_spectra}(a).
The blue circles correspond to the quasi-condensate fraction, $n_{\rm{qc}}$ which decreases smoothly with $T$. We point out that the incoherent fraction $n_\mathrm{inc}/n_0$ is around 7\% at the critical temperature (blue dotted line), for our system parameters~\cite{SM}.
The density of vortices $n_\mathrm{V}$, evaluated from the vortex circulation around closed paths on the numerical grid~\cite{Reeves2014,Billam2014},
is shown in red squares (right axis). 
As expected~\cite{comaron2019quench}, a few (bound) vortices are found at temperatures slightly below $T^{\infty}_{\mathrm{BKT}}$, and $n_\mathrm{V}$ increases steadily with temperature for $T > T^{\infty}_{\mathrm{BKT}}$. 
We show in Sec. S4 of the Supplemental Information \cite{SM} that the value of both $n_\mathrm{qc}$ and $n_\mathrm{V}$  is well converged for our box size, and can thus be interpreted as {\em local} observables.
In \cite{SM} we also report the 
critical exponent $\alpha(T/T^{\infty}_\mathrm{BKT})$,
which accounts for quasi-long-range order.
The {\em non-locality} of this quantity results in logarithmic corrections with the system size, so that the critical value $\alpha=1/4$ is achieved, for the box size used here, for temperatures  $T \simeq 1.25 T^{\infty}_{\mathrm{BKT}}$.
The fact that the impurity spectra discussed below are well converged with the system size (see Sec. S6 of \cite{SM}) confirms the locality of the impurity probe, and its insensitiveness to quasi-long-range order.

In the equal mass case $M\!=\!m$,
and setting $g_\mathrm{BI} = 2 g_\mathrm{BB}$,
the impurity spectral function $S(\omega)$ is shown in  Fig.~\ref{fig:polaron_spectra}(b), as a function of temperature.
At low temperatures, the Bose gas can be approximated as a uniform background of density $n_0$, and the impurity spectrum behaves as a delta-like peak centered at $\omega = g_\mathrm{BI} n_0$.
At higher temperatures,
density fluctuations become important; in particular, vortices are nucleated slightly below $T^{\infty}_{\mathrm{BKT}}$, where  the field develops points of zero density. Conversely,  it seems that not all  density holes correspond to  vortices, as visible from the strong density fluctuations in Fig.~S1 and S2~\cite{SM}.
As a consequence, the impurity can ``bind'' to regions of low density (we recall that the impurity-boson interaction is repulsive). 
Indeed, the green arrow in Fig.~\ref{fig:polaron_spectra}(b) indicates the bound state energy of the impurity within a single vortex, assuming a $T=0$ static bath (see Ref.~\cite{SM} for details).
This binding results in the phenomenology observed in Fig.~\ref{fig:polaron_spectra}(b) at intermediate temperatures, where two spectral lines are reminiscent of the usual repulsive and attractive polaron branches. The attractive branch appears to have a minimum
around $T_\mathrm{BKT}^\infty$,
and  displays a natural broadening (see the spectral slices in Ref.~\cite{SM}).
The blueshift of the attractive branch above $T_\mathrm{BKT}$ is in part due to the increase of the incoherent fraction, which provides a uniform blueshift.
We also remark that a Quantum Monte Carlo study of a repulsive impurity~\cite{pascual2021quasiparticle} identified a redshift of the ground-state energy that increases with $T$.

While these equal mass results seem to connect closely to the typical attractive and repulsive polaron pictures, a very non-trivial behavior is observed when the mass is varied.
For the light impurity of mass $M=0.5 m$
reported in Fig.~\ref{fig:masses}(a), 
a single line dominates the spectrum at all temperatures, which is broadest at intermediate temperatures. As shown in section S7 of ~\cite{SM}, a similar spectrum is obtained  when freezing the bath dynamics,  suggesting that the light impurity is fast enough to cancel out the time dependence of the bath and effectively experience a static bath density.
We then speculate that the presence of a unique line is related to the fact that, in a closed system, the ground state wavefunction can always be chosen real and positive, entailing a sizable zero momentum component;
as a consequence, low lying energy states would carry
most of the oscillator strength.

On the contrary, for the heavier impurity with
$M=2 m$ shown in Fig.~\ref{fig:masses}(b),
one recovers a repulsive line, as in Fig.~\ref{fig:polaron_spectra}(b), but the
 attractive features are quite different.
First of all,  two attractive lines are visible (we refer to the slices at fixed temperatures reported in Fig. S6 in Ref.~\cite{SM} for better visibility), and we relate this to the fact that for this mass ratio the single vortex admits now two $s$-wave impurity bound states, whose energies are indicated by the green arrows (see also Sec. S3 in Ref.~\cite{SM}).
Furthermore, the attractive lines are very weak and broad (notice that the colorscale has been saturated to make these lines more visible). 

In Ref.~\cite{SM} we also calculated the injection spectra for attractive boson-impurity interactions. In this case, only one line is present, broadening with increasing temperatures. The absence of a lower lying peak is probably due to the fact that bumps of high density have a very fast dynamics, and do not support proper bound states, like vortices do in the repulsive case.

~

~

\begin{figure}[t]
\centering
\includegraphics[width=0.99\linewidth]{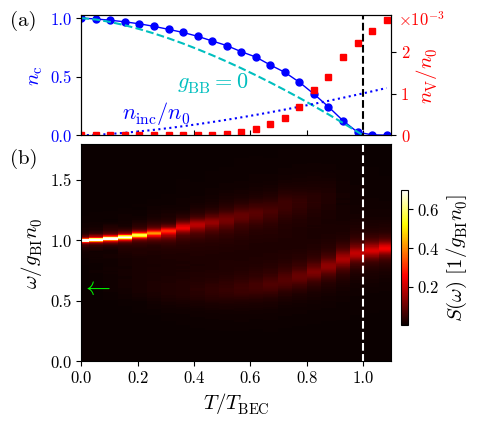} 
\caption{
\textbf{BEC phase diagram and injection spectra for the 3D case.}
For the 3D Bose gas without impurity, we plot
as a function of temperature
the condensate fraction $n_{\rm c}$ (blue circles, left axis),
the incoherent density $n_\mathrm{inc}$ (blue dots) and the density of (projected) vortex ring points (red squares, right axis). 
(b) The injection spectrum $S(\omega)$ of the impurity is plotted as a function of $T$, in the equal mass $M=m$ case. The green arrow indicates the binding energy of the impurity in an isolated vortex at $T=0$. 
The attractive branch $S(\omega)$ displays a minimum close to the threshold for vortex ring nucleation, while above the condensation temperature the gas acts like a uniform thermal gas, so one observes a single Lorentzian peak. 
The colorcode is saturated, since at small temperature $S(\omega)$ consists of a very narrow and high peak.
}
\label{fig:3D}
\end{figure}

\textit{Numerical results in 3D.} It is equally interesting to consider the impurity injection into a 3D Bose gas.
The physics of the Bose gas in the absence of the impurity and for increasing temperature is
illustrated in Fig.~\ref{fig:3D}(a).
First, the condensed fraction 
$n_c =
\llangle |\psi_{\mathbf{k}=0}|^2 \rrangle / n_0
$
 (blue circles, left axis) 
 is lower bounded by the non-interacting, textbook result 
$n_c(g_\mathrm{BB}=0) = 1 - (T/T_{\rm BEC})^{3/2}$, indicated by the cyan dashed line,
and vanishes close to the BEC condensation temperature
 $T_{\rm BEC} \simeq 3.3125 \ n^{2/3}  / m$ ~\cite{pitaevskii2016book}.
The blue dotted line denotes the incoherent fraction.
A very interesting aspect of the 3D Bose gas at finite temperature, which to our knowledge has not been  investigated before, 
is the presence of vortex rings well below the BEC transition temperature.
These rings are analogous to paired vortices, and do not spoil the long-range order. We detect the rings by the application of the 2D vortex detection algorithm, by slicing the 3D space along each of the three directions separately.
The density of vortical points is plotted as red squares in Fig.~\ref{fig:3D}(a) (right axis).
A snapshot of the vortex rings is reported in Fig. S10 of Ref.~\cite{SM}.

The impurity injection spectra, calculated for parameters 
suitable to Rubidium experiments (see Ref.~\cite{SM})
and with $M=m$, 
show  similar features as the 2D ones, with the presence of the attractive line correlated to the nucleation of vortex rings; see Fig.~\ref{fig:3D}(b).
At the BEC transition, however, the vanishing of the condensate fraction entails that the fluid behaves like a uniform thermal gas. As a consequence, the spectrum features a single peak around $\omega \sim g n_0$.

~

\textit{Conclusions and discussion.}
In summary, we have computed the injection spectra of an impurity in a weakly-interacting Bose gas at finite temperature, using stochastic  classical-field methods and for repulsive impurity-boson interactions.
The binding of the impurity to vortices and density holes makes the spectra sensitive to the BKT transition in 2D, while in 3D the vortex nucleation and the condensation transitions occur at substantially different  temperatures. 
Our numerics has been performed for parameters suitable to Rubidium ultracold gases, and is accessible in current experiments.

Our work opens several intriguing directions. First, it suggests that it may be possible to obtain  indirect signatures of the non-local BKT transition via a local probe, which may enable the detection of superfluidity in TMD  EXI experiments. A more detailed modeling, keeping into account the fermionic and dipolar  nature of the exciton fluids, will be pursued in the future.

Also, our results bear strong implications for the more conventional Bose polaron setup, i.e.~with {\em attractive} impurity-boson interactions, for which no reliable spectral predictions are available beyond small temperatures.
Here, we expect that the repulsive branch will be effectively described by the present formalism, and will feature a splitting at finite temperatures. This claim (as well as modifications in the genuinely attractive branch)
should be investigated in future work. A promising strategy would be to apply the finite temperature Lee-Low-Pines transformation~\cite{dzsotjan2020dynamical}
together with the GPE generalization to this frame~\cite{drescher2020theory}.
We estimate this program to be very expensive computationally, and we leave it for future research.

Finally, it would also be interesting to explore the {\em nonlinear}
polaron spectroscopy regime, by considering the recoil of the bosons and its effect on the superfluid transition, which is relevant at a finite density of impurities, as well as introducing interactions between the impurities.
In this case, ejection spectra could also provide a rich framework.
Another straightforward prospective concerns driven-dissipative polariton fluids, where the onset of BKT physics can be imaged on the micrometer scale~\cite{2018caputo,Comaron_2021,comaron2024coherence}.

\textit{Acknowledgements.}
We acknowledge inspiring discussions with Jean Dalibard, Eugene Demler, Franco Dalfovo, Nick Proukakis, Michiel Wouters, Marzena Szymanska, Davide Galli, Georg Bruun, Dario Ballarini, Daniele Sanvitto, Artem Volosniev, and Haydn Adlong.
Work in Brussels is financially supported by the ERC grant LATIS, the EOS project CHEQS and the FRS-FNRS (Belgium). 
P.C was supported by ``Quantum Optical Networks based on Exciton-polaritons'' (Q-ONE, N. 101115575, HORIZON-EIC-2022-PATHFINDER CHALLENGES EU project), ``Neuromorphic Polariton Accelerator'' (PolArt, N.101130304, Horizon-EIC-2023-Pathfinder Open EU project), ``National Quantum Science and Technology Institute'' (NQSTI, N. PE0000023, PNRR MUR project), ``Integrated Infrastructure Initiative in Photonic and Quantum Sciences'' (I-PHOQS, N. IR0000016, PNRR MUR project).
Views and opinions expressed are however those of the author(s) only and do not necessarily reflect those of the European Union or European Innovation Council and SMEs Executive Agency (EISMEA). Neither the European Union nor the granting authority can be held responsible for them. P.C. was also supported from the Engineering and Physical Science Research Council: Grants No. EP/V026496/1, No. EP/S019669/1, and No. EP/S021582/1.
A.I was supported by the Swiss National Science Foundation (SNSF) under Grant Number 200021-204076.
Computational resources have been provided by the Consortium des \'Equipements de Calcul Intensif (C\'ECI), funded by the Fonds de la Recherche Scientifique de Belgique (F.R.S.-FNRS) under Grant No. 2.5020.11 and by the Walloon Region.

\bibliography{bibliography}

~

\newpage

\makeatletter
\renewcommand \thesection{S\@arabic\c@section}
\renewcommand \thefigure{S\@arabic\c@figure}
\makeatother

\newpage

~

\newpage

\title{Supplemental Information for  ``Quantum impurities in finite-temperature Bose gases: \\Detecting vortex proliferation across the BKT and BEC transitions''}





\maketitle

\onecolumngrid

\section{Physical parameters}

First of all, we recall that the standard (noiseless and unprojected) Gross-Pitaevskii equation  can be put in adimensional form by measuring length in units of the healing length $\xi = (2m g_{ \rm BB} n_0)^{-1/2}$ and frequency in units of $g_{ \rm BB} n_0$ (having set $\hbar = 1$).
In 2D and at finite temperature, the product $m g_{ \rm BB}$ also matters (e.g. see the expression for $T_{\rm BKT}$),
and in the SPGPE we have the (adimensional) relaxation coefficient $\gamma$.
Given a box of area $V=L^2$, with $L$ its linear extension (we recall \textit{en passant} that we use periodic boundary conditions), the only irreducible parameters  to be set are then
$L/\xi$, $m g_{ \rm BB}$, $T/g n_0$ (or, as we use in the main text, $T/T^\infty_{\text{BKT}}$), and $\gamma$.

To connect with cold atom experiments,
we use values that have been previously used in Ref.~\cite{comaron2019quench}
to model a 2D $^{87}$Rb gas at Laboratoire Kastler Brossel~\cite{Chomaz2015,ville2018,saintjalm2019}.
These are a mass
$m=1.4431 \cdot 10^{-25}$kg, average density $n_0 = 9 \mu$m$^{-2}$, and interaction strength 
$g_{ \rm BB} \simeq 0.53$ nK
$\cdot \mu$m$^{2}$, which yields the crucial parameter
$g_{ \rm BB} m  = 9.5 \cdot 10^{-2}$.
(The Boltzmann constant has also been set to 1, and we report energies and lengths in nK and $\mu$m, respectively.)
Given that $\xi \simeq 0.76 \mu$m,  for the main results we use a box size of $L=20 \mu$m (some finite size comparison is reported below, in Sec. S4), while we scan the temperature across $T^\infty_{\text{BKT}} \simeq 32$ nK.
The results are weakly dependent on $\gamma$, which is set to $\gamma=0.05$. Finally, we set $g_{ \rm BI}=2 g_{ \rm BB}$ for the impurity-boson interaction, but we expect that this value can be broadly tuned via Feshbach resonances.

In 3D, we use $n_0 = 12 \mu$m$^{-3}$ and $g_{ \rm BB} \simeq 0.5$ nK
$\cdot \mu$m$^{3}$.
This corresponds to 
$g_{ \rm BB} = {4\pi a_{ \rm 3D}\hbar^2 }/{m}$, with
a 3D scattering length
$a_{ \rm 3D}=135 a_{ \rm Bohr}$, having denoted $a_{ \rm Bohr}$ the Bohr radius;
for comparison, the background scattering length has been measured to be 
around $100 a_{ \rm Bohr}$~\cite{Buggle2004}. In any case, the Berezinskii-Kosterlitz-Thouless (BKT) physics depend only logarithmically on $g_{ \rm BB}$.
The resulting (non-interacting) condensation temperature is $T_{\rm BEC} \simeq 97$ nK and the healing length $\xi \simeq 0.68 \mu$m.

~

~

~

\section{Numerical implementation}

A pseudo-spectral 4th-order Runge-Kutta approach is used to evolve the SPGPE.
The vortex locations were detected using the open-source Julia library \texttt{VortexDistributions.jl}, developed by Ashton S. Bradley~\cite{Reeves2014,Billam2014}.

~

~

~

\section{Typical density and phase patterns}

We report random snapshots of the density and phase of the Bose field, at two temperatures, slightly below
and above the BKT transition.

\begin{figure*}[h]
\centering
\includegraphics[width=0.99\textwidth]{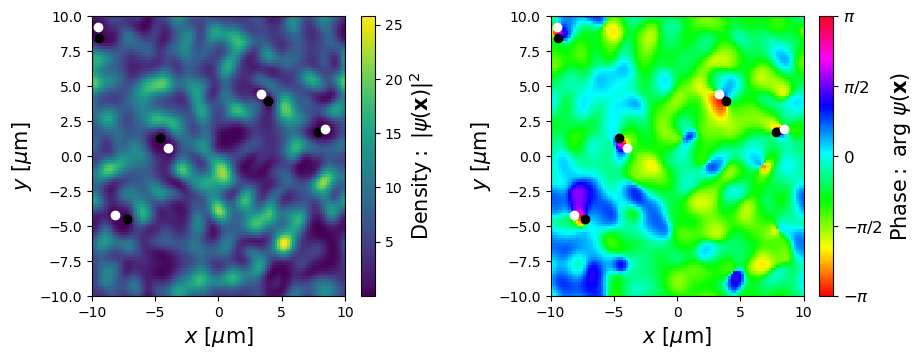} 
\caption{
Typical snapshot of the density (left) and phase (right)
of the quasi-condensate at $T \simeq 0.94 T_{\rm BKT}$. 
The black and white dots denote vortices and antivortices, respectively.
One can notice a few bound vortices, as well as density deeps which do not correspond to any vortex.
The topological defects are detected using the open-source Julia library \texttt{VortexDistributions.jl}, developed by Ashton S. Bradley~\cite{Reeves2014,Billam2014}.
}
\label{fig:typical_nphi_below}
\end{figure*}

\begin{figure*}[h]
\centering
\includegraphics[width=0.99\textwidth]{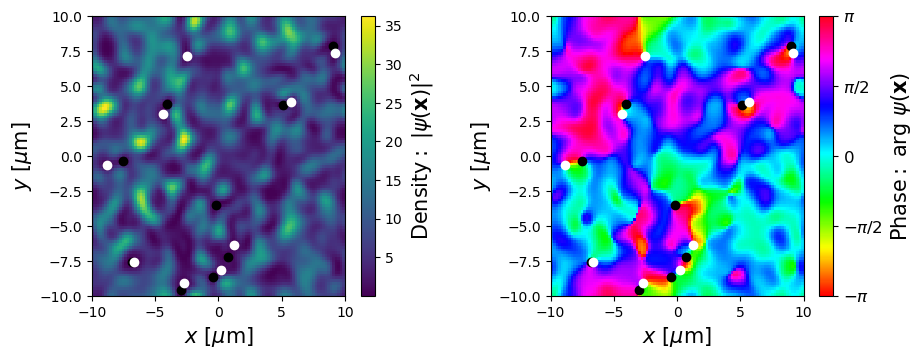} 
\caption{
Typical snapshot of the density (left) and phase (right)
of the quasi-condensate at $T \simeq 1.25 T_{\rm BKT}$. 
The black and white dots denote vortices and antivortices, respectively.
Apart from  a few bound vortex-antivortex pairs, one can spot two unbound vortices, with the phase performing a full $\pm 2\pi$ winding over a large distance.
}
\label{fig:typical_nphi_above}
\end{figure*}

~

~

\section{Quasi-long-range order and finite size effects}

\begin{figure*}
\centering
\includegraphics[width=0.99\textwidth]{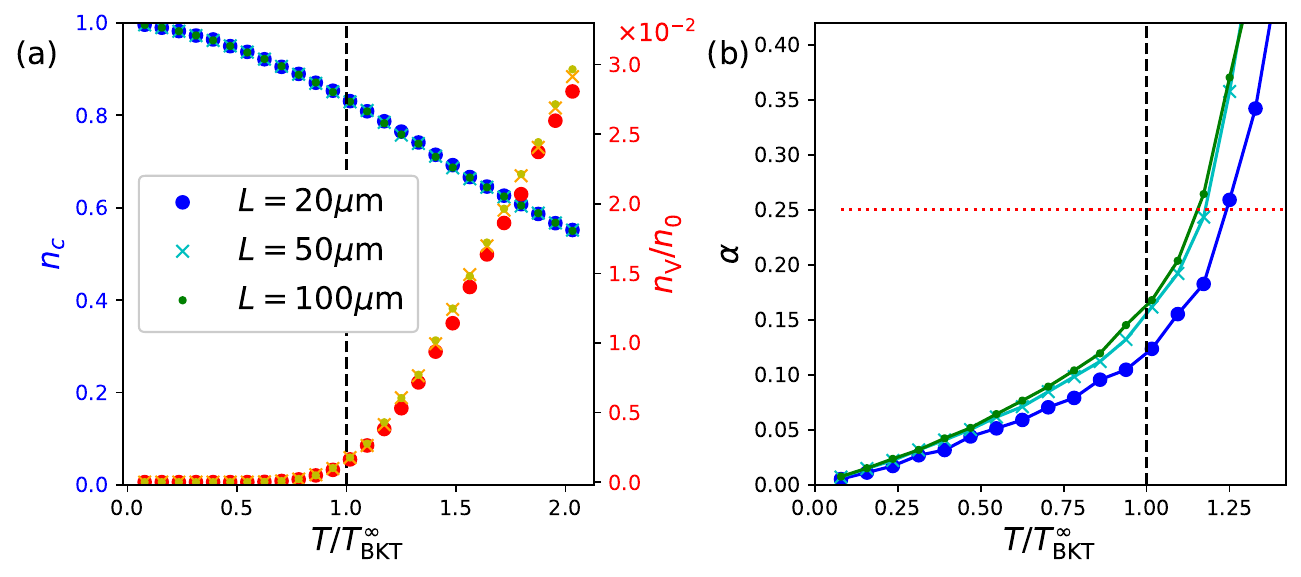} 
\caption{
(a)
The quasi-condensate density $n_{ \rm qc}$ (cold colors, left axis) and vortex density $n_{ \rm v}$ (warm colors, right axis) calculated at different temperatures of the 2D Bose gas.
Different symbols (circle, cross, dot) correspond to different box linear sizes
($L=20,50,100 \mu$m, respectively).
(b) Exponent $\alpha$ extracted from an  algebraic fit of the $g^{(1)}(r)$ correlator.
The critical value $\alpha_c = 0.25$ predicted by BKT theory in the thermodynamic limit is shown as a horizontal dotted red line.
}
\label{fig:fitting_results2}
\end{figure*}

In this section we characterize the power-law decay of spatial fluctuations in our 2D simulations and the effect of the system size  on the extension of the critical region for the BKT phase transition.
In a two-dimensional system experiencing a BKT phase transition, it is anticipated that: (i) for temperatures above $T_{\mathrm{BKT}}$, 
superfluidity is absent,
and the first-order correlation function follows an exponential decay $g^{(1)}(r) \sim e^{-r / \ell}$, where $\ell$ stands for the correlation length; (ii) for temperatures below $T_{\mathrm{BKT}}$, superfluid
behavior 
manifests itself in an algebraic decay of
the correlation function,
$g^{(1)}(r) \sim r^{-\alpha}$, with $\alpha(T)$ being a temperature dependent exponent.

To characterize the equilibrium state of our finite-size uniform 2D Bose gas in the absence of impurities, we compute the first order correlation function $g^{(1)}(r)$ by simulating the SPGPE of Eq. (3) of the main text and averaging over many stochastic realizations.
Once this function is computed, we fit it with the function
\begin{equation} 
g_{\text{alg}}^{(1)}(r) \propto r^{-\alpha}.
\end{equation}

In Fig.~\ref{fig:fitting_results2}.(a)
we report
the quasi-condensate density $n_{ \rm qc}$  and vortex density $n_{ \rm v}$
for different system sizes.
Since these quantities are local observables, the finite size effects are negligible.
On the contrary, the 
superfluid behavior is a quasi-long-range effect, and is more sensitive to the size of the system.
In Fig.~\ref{fig:fitting_results2}.(b), we plot the  algebraic exponent $\alpha(T)$ extracted from the fit,
for three different sizes of the numerical box
($L=20 \mu$m corresponds to the size used for the results reported in the main text.). 
%
The critical exponent $\alpha \approx 0.25$, expected at $T^{\infty}_{\mathrm{BKT}}$, is  found at $T/T^{\infty}_{\mathrm{BKT}} \approx {1.25}$ for $L=20 \mu$m,
and at $T/T^{\infty}_{\mathrm{BKT}} \approx {1.15}$ for $L=100 \mu$m. 
We attribute this shift to the effect of the finite size of the system. 
These findings align with the expectation that the critical region diminishes in size as it approaches the thermodynamic limit~\cite{Foster2010,Gawryluk2019}.
We remark that, according the Kosterlitz-Thouless renormalization group analysis, the scaling of the critical point is logarithmic in the system size, so locating the phase transition with high precision requires exponentially large $L$~\cite{Prokofev2002}.
 The shift of the critical exponent to a temperature greater than $T^{\infty}_{\mathrm{BKT}}$ has also been observed in Ref.~\cite{Foster2010,Gawryluk2019,comaron2019quench},
 where the existence of an intermediate critical crossover region has also been studied in detail, and its orgin identified in the finite size of the system.

~

~

\section{Vortex-impurity binding energy}

We evaluate here the binding energy of an impurity bound to an ideal vortex, i.e. the stationary soliton occurring at zero temperature, originally obtained by Pitaevskii~\cite{pitaevskii1961}.
As demonstrated in \cite{Berloff2004}, the vortex radial profile is excellently approximated by a Pad\'e expansion.
Namely, the density in the vortical configuration reads in polar coordinates $n(r)=n_0 f_{\rm Berloff}(r/\xi)$,
where
$ f_{\rm Berloff}(s)=(0.3437 \cdot s^2 + 0.0286 \cdot s^4) / (1 + 0.3333 \cdot s^2 + 0.0286 \cdot s^4)$.

\begin{figure*}[h]
\centering
\includegraphics[width=0.55\textwidth]{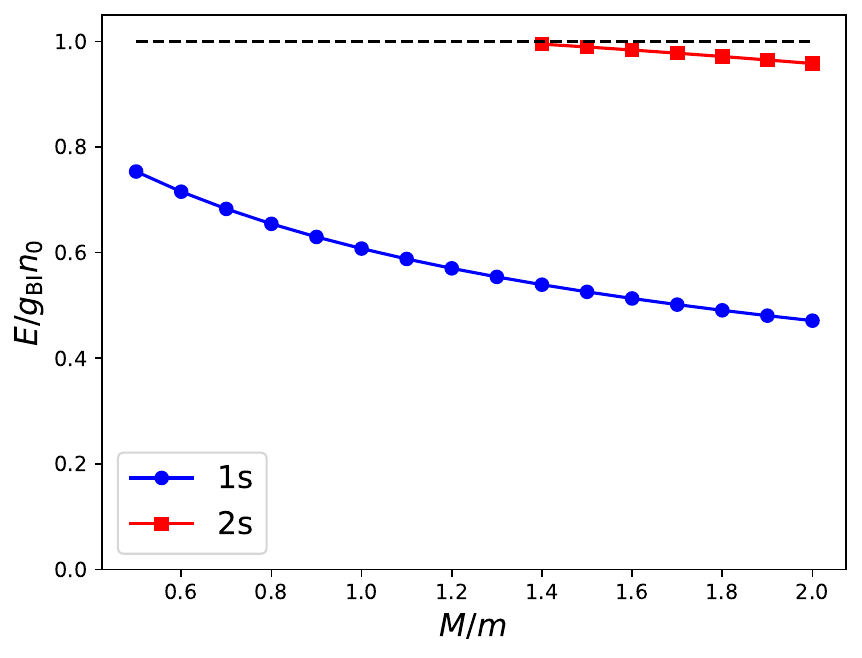} 
\caption{
Energy of the impurity bound to an ideal vortex, for different impurity mass (and assuming $g_{\rm BI} = 2 g_{\rm BB}$).
The $1s$ bound state (blue circles) exists for any value of $M$, as expected in 2D. 
The $2s$ bound state (red squares), appear instead at large enough mass.
}
\label{fig:Eb}
\end{figure*}

We then solve for the radially symmetric bound states of the impurity (higher angular momentum states are dark), by means of exact diagonalization on a spatial grid.
The results are reported below, in Fig.~\ref{fig:Eb}.

\section{Spectral slices}

~

\begin{figure*}[h]
\centering
\includegraphics[width=0.49\textwidth]{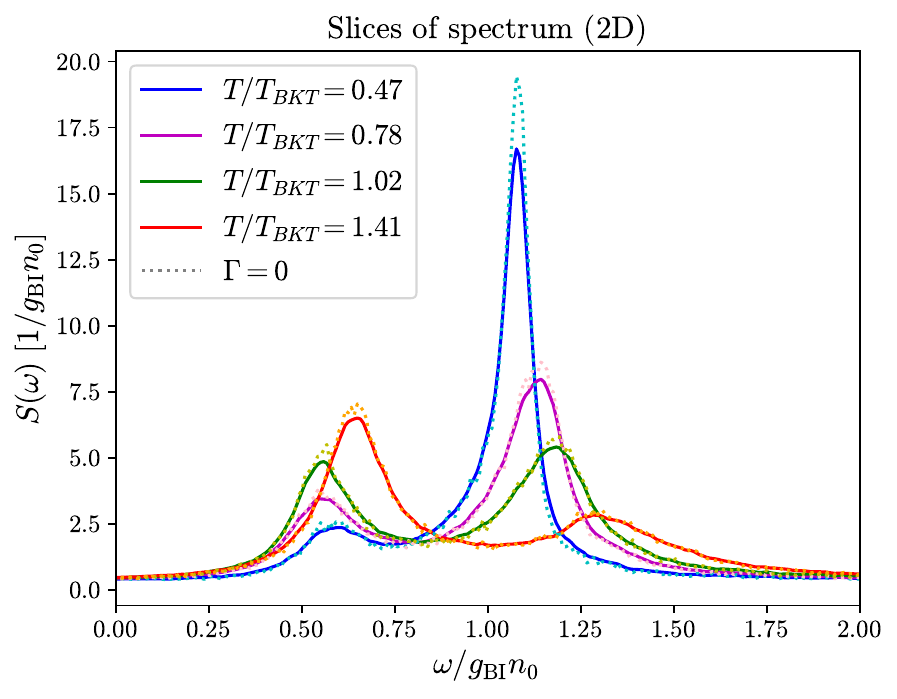} 
\includegraphics[width=0.49\textwidth]{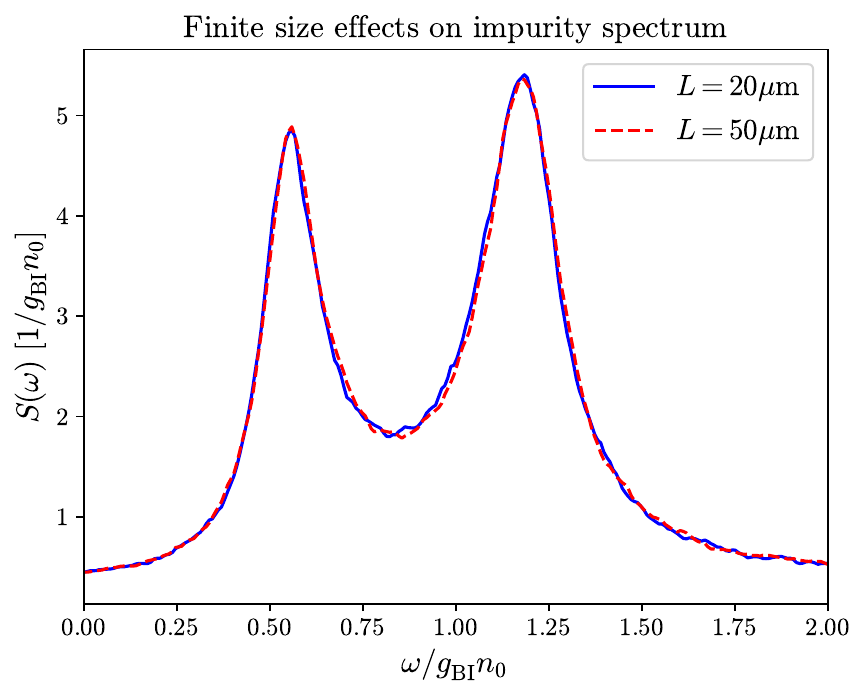} 
\caption{
(left) We report four slices of the  spectrum of Fig. 2.(b), i.e. the two-dimensional system with equal masses, corresponding to four different temperatures.
The solid lines denote the spectrum reported in the main text, artificially smoothed by convolving with a Lorentzian of width $\sim 0.01 g_{\rm BI} n_0$ (which is necessary for the graphical rendering at vanishing temperatures). The dotted lines correspond to the case with no artificial broadening, which are slightly noisy. This plot shows that both the attractive and repulsive branches display a naturally broadened linewidth; this is in contrast with standard attractive polarons, which are  quasi-particles with delta-like peaks. Also, at relevant temperatures, the artificial broadening is inconsequential.
(b)
The spectra reported in the main text are converged with respect to the size of the system. This is shown in this panel, where the spectrum at $T=1.41 T^\infty_{ \rm BKT}$ is computed for two box sizes, $L=20\mu$m (as in the main text, blue solid line) and 
$L=50\mu$m (red dashed line). The convergence is basically perfect.
}
\label{fig:3slice}
\end{figure*}

\begin{figure*}[h]
\centering
\includegraphics[width=0.54\textwidth]{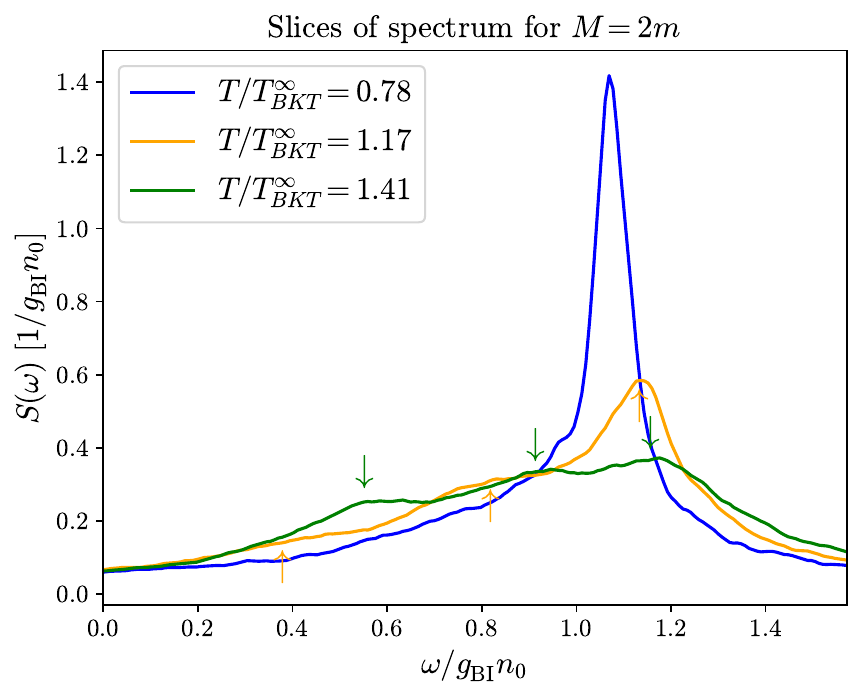} 
\caption{
We report three slices of the $M=2m$ spectrum of Fig. 2.(b), corresponding to three different temperatures.
The presence of three peaks for $T \gtrsim T_{\rm BKT}$ is highlighted by the arrows.
}
\label{fig:3slice}
\end{figure*}

\begin{figure*}[h]
\centering
\includegraphics[width=0.54\textwidth]{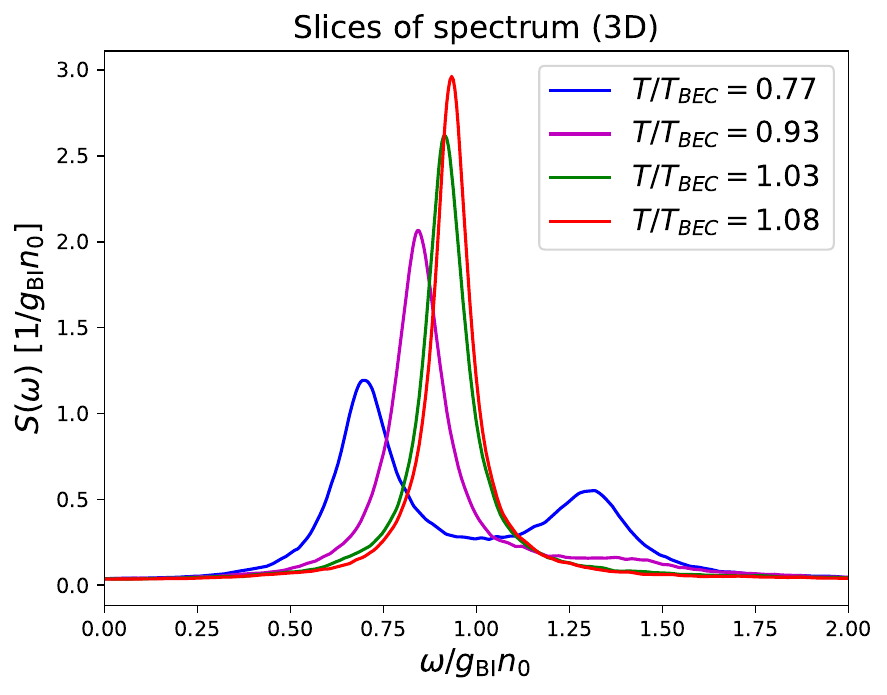} 
\caption{
We report four slices of the  spectrum of Fig. 3.(b), i.e. the three-dimensional system, corresponding to four different temperatures close to the condensation transition.
Below $T_{ \rm BEC}$ a bimodal structure is clearly visible, which is absent above transition, where a single Lorentzian peak becomes narrower for increasing temperatures. Remarkably, this change of behavior occurs on a small temperature range. 
}
\label{fig:slice_3D}
\end{figure*}

\newpage

\section{Frozen bath dynamics}

In the main text we presented the injection spectrum of the light $M=m/2$ impurity, and remarked that it features only one clear branch, on the attractive side.

\begin{figure*}[h]
\centering
\includegraphics[width=0.54\textwidth]{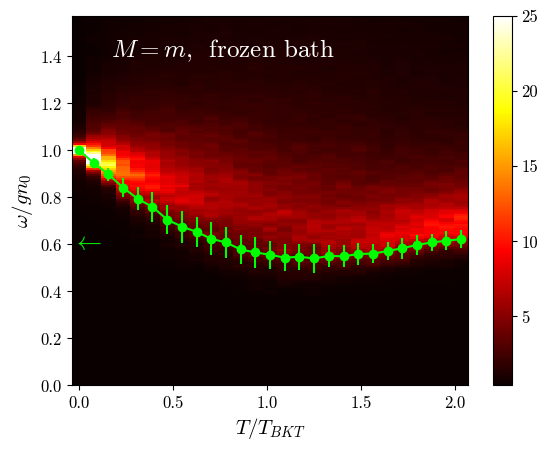} 
\caption{
We report the spectrum of the impurity injected in the Bose bath, assuming that the dynamics of the bosons is frozen in the moment the impurity is injected. The green circles correspond to the average ground state energy of the impurity over different stochastic realizations of the bath, while the bars denote the standard deviation.
}
\label{fig:frozen}
\end{figure*}

Here, we argue that this result can be qualitatively understood by observing that a light impurity has a fast dynamics, and therefore the dynamics of the bath looks slower to the impurity.
We corroborate this perspective by calculating the spectrum of the impurity, injected into a {\em static} stochastic background, obtained by freezing the dynamics of the bath.
The spectrum is averaged over $N_s=100$
thermal realizations of the bath.
This is shown in Fig.~\ref{fig:frozen}.
Notice that we consider here the $M=m$ impurity, since we want to highlight the striking qualitative differences with respect to the fully dynamic calculation of Fig. 1.(b).
For each stochastic realization, we also compute the ground state energy of the impurity, which makes sense because now the bath plays the role of a static, conservative potential field.
The average ground state energy is shown as green circles, with the bars denoting the standard deviation (which depends on the size of the simulation box).

Interestingly, the spectrum mainly features a single broad branch, peaking on the attractive side at energies comparable with the ground state.
We relate this fact to Feynman's observation that the ground state of a particle in the absence of magnetic fields is always real and can be chosen to be positive. 
This entails that the ground state always  has a finite oscillator strength. It then does make sense that other low lying states concentrates a large part of the oscillator strength.
It would be interesting to study injection spectroscopy into disordered potentials in more general settings.

~

\section{Attractive boson-impurity interactions}

\begin{figure*}[h]
\centering
\includegraphics[width=0.99\textwidth]{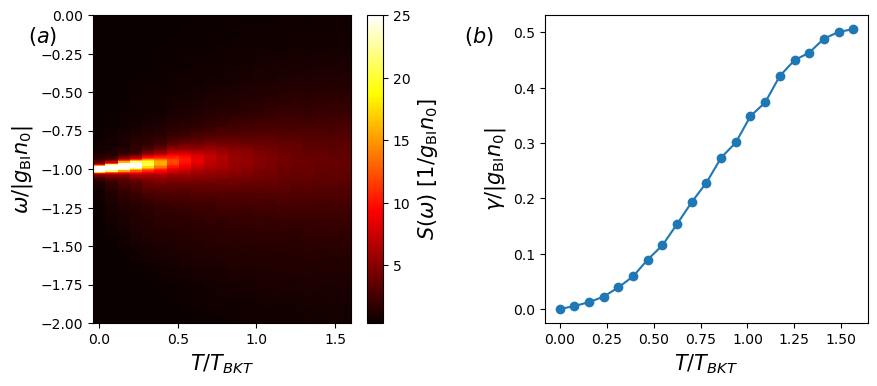} 
\caption{
(a) Injection spectrum for attractive boson-impurity interactions. (b) Half-width-at-half-maximum of the peak of panel (a).
}
\label{fig:attractive}
\end{figure*}

In this paragraph, we calculate the injection spectrum for $g_{\rm BI}=-2g_{\rm BB} < 0$.
The spectrum is reported in Fig. \ref{fig:attractive}(a).
One can see that there is only one line, which broadens with temperature, with half-width-at-half-maximum 
$\gamma$ reported in panel (b).
The presence of a single peak illustrates the statement of the main text, that neglecting quantum two-body correlations is a good approximation only in the $g_{\rm BI} > 0$ repulsive case. Indeed, while the present spectrum could provide a good indication in the limit of small attractive $g_{\rm BI}$, where mean-field theory works, at finite $g_{\rm BI} < 0$ we fully miss the repulsive branch which characterize Bose polaron spectra. For repulsive interactions, genuine 2-body bound states are not supported and no such strong correlations are present,  and one can rely on our stochastic mean-field 
description.

In other words, if a fully quantum calculation were possible, we would expect the following temperature effects (for $g_{\rm BI} < 0$). First, the genuine AP peak would be broadened. This has been computed here in the mean-field regime, and in the limit $g_{\rm BI} \to - \infty$ one can think in terms of a brownian motion of a molecular state.
Moreover, at positive energies the scattering of the impurity with the bosons is effectively captured by a 
repulsive coupling.
Then, we expect the repulsive polaron to further split, according to the description provided in the main text. Since the broadening of the RP is a higher-order effect in the effective coupling constant, our splitting (of a linear order) should be visible for a deep enough two-body bound state.
In conclusion, we expect our results to have an impact on the finite temperature Bose polaron spectra also in the standard 
$g_{\rm BI} < 0$ framework.

\newpage

\section{Snapshot of 3D vortex rings}

\begin{figure*}[h]
\centering
\includegraphics[width=0.52\textwidth]{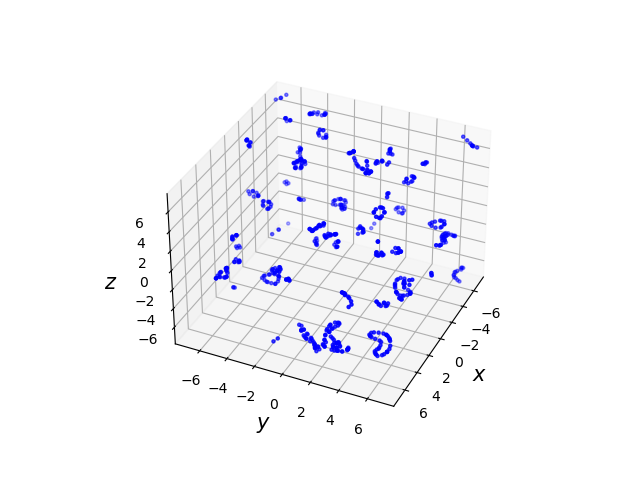} 
\caption{
A snapshot of the vortex rings present in a particular realization of the 3D Bose gas at $T=0.55 T_{\rm BEC}$.
Namely, we slice the field configuration over each of the three directions, and we apply the 2D vortex detection algorithm from the library \texttt{VortexDistributions.jl}~\cite{Reeves2014,Billam2014}. All the points found are plotted, revealing the presence of vortex rings, which, being topologically closed strings, do not spoil the long-range coherence of the condensate.
}
\label{fig:vrings}
\end{figure*}




\end{document}